\documentclass[aps,prl,twocolumn,nofootinbib,superscriptaddress]{revtex4-1} 

\usepackage{epsfig}
\usepackage{color}
\usepackage[latin1]{inputenc}
\usepackage{float,amsmath}
\usepackage{graphicx}

\begin{document}

\title{Pseudorapidity correlations in heavy ion collisions from viscous fluid dynamics}

\author{Akihiko Monnai}
\affiliation{RIKEN BNL Research Center, Brookhaven National Laboratory, Upton, NY 11973, USA}

\author{Bj\"orn Schenke}
\affiliation{Physics Department, Brookhaven National Laboratory, Upton, NY 11973, USA}

\begin{abstract}
We demonstrate by explicit calculations in 3+1 dimensional viscous relativistic fluid dynamics how two-particle pseudorapidity correlation functions in heavy ion collisions at the LHC and RHIC depend on the number of particle producing sources and the transport properties of the produced medium.
In particular, we present results for the Legendre coefficients of the two-particle pseudorapidity correlation function, $a_{n,m}$, in Pb+Pb collisions at 2760 GeV and Au+Au collisions at 200 GeV from viscous hydrodynamics with three dimensionally fluctuating initial conditions.
Our results suggest that the $a_{n,m}$ provide important constraints on initial state fluctuations and the transport properties of the quark gluon plasma.
\end{abstract}

\maketitle


\section{Introduction}
The study of the Fourier coefficients $v_n$ of the azimuthal particle distribution in heavy ion collisions at the Relativistic Heavy Ion Collider (RHIC) and the Large Hadron Collider (LHC) has led to great insight into the initial state fluctuations and hydrodynamic evolution of the produced system \cite{Gale:2013da}. In particular, it has led to the conclusion that the system has a very small viscosity, close to the lower bound conjectured using AdS/CFT correspondence \cite{Kovtun:2004de}.

Recently, the ATLAS collaboration has presented results on the expansion of two-particle pseudorapidity correlations into Legendre polynomials \cite{ATLAS-CONF-2015-020}. The obtained coefficients contain important information on the fluctuations of the particle multiplicity in pseudorapidity.

In this letter we explore how two-particle pseudorapidity correlations from hydrodynamic calculations can give insight into the number of sources for particle production and their correlations, as well as the shear and bulk viscosity of the system.
To achieve this, we introduce a simple initial state model that extends the conventional Monte Carlo (MC) Glauber model \cite{Miller:2007ri} to include fluctuations of valence quarks in rapidity and thus produces three dimensional fluctuating distributions of net baryon number and entropy density. We then use this model to generate the input for 3+1 dimensional viscous hydrodynamic calculations and compute rapidity distributions of charged hadrons and two-particle rapidity correlations. We then analyze the effect of i) the number of sources and ii) the shear and bulk viscosity of the system on the Legendre coefficients.

\section{Initial state model and hydrodynamic evolution}
\label{sec:simpleGlauber}
Fluctuations in rapidity have been included in hydrodynamic calculations via UrQMD \cite{Petersen:2008dd,Karpenko:2015xea}, EPOS \cite{Werner:2010aa}, or AMPT \cite{Pang:2014pxa} initial conditions.
To study the effect of fluctuations in rapidity in addition to fluctuations in the transverse plane without being dependent on a complicated string with many parameters, we introduce a simple model that is a straight forward extension to the MC Glauber model to include longitudinal fluctuations.
In particular, we will employ a Monte Carlo implementation of the Lexus model \cite{Jeon:1997bp} to provide a simple parametrization of the rapidity distribution of wounded nucleons (or constituent quarks), which is based on experimental observations in proton-proton collisions.

\emph{3DMC-Glauber model}
For each nucleus we sample the three-dimensional spatial distribution of nucleons from a Woods-Saxon distribution.
We then sample valence quark positions within each nucleon from an exponential distribution.
After overlaying the two nuclei in the transverse plane, separated by the sampled impact parameter $b$, wounded
quarks are determined using the quark-quark inelastic cross-section $\sigma_{qq}$, which can be obtained geometrically by requiring that the experimentally determined nucleon-nucleon cross section is recovered.
We do, however, treat the quark-quark cross section in heavy ion collisions as an independent, free parameter. When we choose $\sigma_{qq}=3\,{\rm mb}$ at $\sqrt{s}=200\,{\rm GeV}$, we can reproduce the experimental multiplicity and pseudorapidity distribution of charged hadrons without including additional negative binomial fluctuations. We employ a Gaussian wounding where quarks are determined to be participants with a Gaussian probability \cite{Bialas:2006qf,Broniowski:2007nz}. 

As mentioned above, to obtain the longitudinal distribution of the initial baryon number we employ the Lexus model \cite{Jeon:1997bp}. Here, the idea is that the rapidity distribution of nucleons in heavy ion collisions can be obtained by linear extrapolation from the distribution in p+p collisions. That distribution is parametrized and fit to experimental data. We use this model for valence quarks such that the probability for a quark with rapidity $y_P$ to end up with rapidity $y$ after a collision with a quark with rapidity $y_T$ (from the other nucleus) is given by \cite{Jeon:1997bp}:
\begin{align}\label{eq:Q}
  Q(y-y_T,&y_P-y_T,y-y_P) = \nonumber\\
&\lambda \frac{\cosh(y-y_T)}{\sinh(y_P-y_T)} + (1-\lambda) \delta(y-y_P)\,,
\end{align}
which corresponds to a flat distribution in longitudinal momentum $p_L$. In the original work \cite{Jeon:1997bp} $\lambda$ is the fraction of nucleon-nucleon scatterings that result in a hard collision. Here, we treat $\lambda$ as a free parameter, regulating the stopping power of the collision. It can be fixed by fitting the net baryon distribution to experimental data. Generally, we find a good fit to the net baryon rapidity distribution when $\sigma_{qq} \lambda \approx 2\,{\rm mb}$.

The initial rapidities are distributed according to the quarks' $x$ value, which is determined by the valence quark parton distribution functions (PDFs).\footnote{The $x$ values are sampled from CT10 NNLO parton distribution functions \cite{Gao:2013xoa} at $Q^2=1\,{\rm GeV^2}$ with EPS09 nuclear correction \cite{Eskola:2009uj} using LHAPDF 6.1.4 \cite{Buckley:2014ana}.} 
Initial rapidities are thus $
  y = \pm {\rm ln} (x \sqrt{s}/2m_q)\,,$
with the sign depending on whether it is a quark in the projectile (+) or the target (-).
We employ a valence quark mass of $m_q=0.34\,{\rm GeV}$ and assume zero transverse momentum initially.

To systematically organize the collisions of all quarks, we have quark pairs collide subsequently with increasing inter-quark distance $\Delta z$.
We then convert rapidity to space-time rapidity $\eta_s$ to obtain a three dimensional event-by-event distribution of quarks. To assign a baryon density, this distribution needs to be smeared and we do this by introducing Gaussians in the transverse plane with width $\sigma_T = 0.4\,{\rm fm}$, and width in space-time rapidity of $\sigma=0.2$.

Next we determine the entropy density distribution. We deposit a fixed entropy between the rapidities of each wounded quark and its last collision partner and smear the edges with half Gaussians of the same width as used for the baryon density distribution. This leads to the following form of the rapidity dependence of the entropy density per wounded quark pair
\begin{align}
  s(y, &y_P,y_T)= \mathcal{N}\exp[- \theta(-y + \min(y_P, y_T))\nonumber \\ &~~~~~~~~~~~~~~~~~~~~~~\times (y - \min(y_P, y_T))^2/2\sigma^2\nonumber\\ &~~~~~~~~~~~~~~~~~~~~~~- \theta(y - \max(y_P, y_T))\nonumber\\ &~~~~~~~~~~~~~~~~~~~~~~\times(y - \max(y_P, y_T))^2/2\sigma^2]\nonumber \\ &\times \left(\max(y_P, y_T) - \min(y_P, y_T) + \sqrt{2 \pi}\sigma\right)^{-1}\,,
\end{align}
where $y_T$ and $y_P$ are the rapidities of the colliding quarks, and $\mathcal{N}$ determines the normalization, which is the same for every ``tube'' and adjusted to fit the experimental data.

In the transverse plane, we smear the entropy density around the center of mass position of each pair by a Gaussian of width $\sigma_T=0.4\,{\rm fm}$.

This way of initializing the entropy density leads to the correct centrality dependence of the total multiplicity. We note that energy and momentum conservation is not explicitly fulfilled, however, we are not including any transverse momentum production or very high momentum partons, which will carry away some of the energy and momentum of the collision and are not part of the bulk medium we are describing.

A different initial state model using random rapidities for wounded nucleons was employed in \cite{Bozek:2015bna}.

\emph{Hydrodynamics}
We integrate the above initial condition into the 3+1 dimensional viscous relativistic fluid dynamic simulation \textsc{Music} \cite{Schenke:2010nt,Schenke:2010rr,Schenke:2011bn,Gale:2012rq}. In addition to numerically solving the equations for the conservation of energy and momentum
$\partial_\mu T^{\mu\nu} = 0\,,$ and the baryon current $\partial_\mu J_B^\mu = 0\,,$
we solve the relaxation-type equations derived from kinetic theory \cite{Denicol:2012cn,Denicol:2014vaa}
\begin{align}
  \tau_\Pi \dot{\Pi} + \Pi &= -\zeta \theta - \delta_{\Pi\Pi} \Pi \theta + \lambda_{\Pi\pi} \pi^{\mu\nu}\sigma_{\mu\nu}\\
  \tau_\pi \dot{\pi}{\langle \mu\nu \rangle} + \pi^{\mu\nu}  &= 2 \eta \sigma^{\mu\nu} - \delta_{\pi\pi} \pi^{\mu\nu}\theta + \phi_7 \pi_\alpha^{\langle \mu} \pi^{\nu\rangle \alpha} \nonumber\\ 
  &~~~~-\tau_{\pi\pi} \pi_\alpha^{\langle\mu} \sigma^{\nu\rangle \alpha} + \lambda_{\pi \Pi}\Pi \sigma^{\mu\nu}\,.
\end{align}
The transport coefficients $\tau_\Pi$, $\delta_{\Pi\Pi}$, $\lambda_{\Pi\pi}$, $\tau_\pi$, $\delta_{\pi\pi}$, $\phi_7$, $\tau_{\pi\pi}$, and $\lambda_{\pi\Pi}$ are fixed using formulas derived from the Boltzmann equation near the conformal limit \cite{Denicol:2014vaa}.
At zero baryon chemical potential the ratio $\eta/s$ will be chosen to be constant in this work, and the temperature dependence of the ratio of bulk viscosity to entropy density $\zeta/s$ is parametrized as in \cite{Denicol:2009am,Ryu:2015vwa}, except that we gradually reduce the constant value at low temperature such that effects of the bulk $\delta f$ corrections \cite{Monnai:2010qp} are kept to a minimum. 
Because we include finite baryon chemical potential $\mu_B > 0$, we replace $s$ in the above expressions by $(\varepsilon + P)/T$, motivated by the relativistic limit of the fluidity measure introduced in \cite{Liao:2009gb}.

The equation of state, which includes finite baryon chemical potential, is constructed by interpolating the pressures of hadronic resonance gas and lattice QCD \cite{Borsanyi:2013bia,Borsanyi:2011sw}.

We leave a detailed description of the initial state model and the newly constructed equation of state for a longer paper in the future.

After Cooper-Frye freeze out at an energy density of $0.1\,{\rm GeV}/{\rm fm}^3$, the calculation of thermal spectra including $\delta f$ corrections \cite{Monnai:2010qp} for shear and bulk viscosities, and resonance decays, we obtain the final hadron spectra as functions of transverse momentum $p_T$ and pseudo-rapidity $\eta_p$.

\section{Two particle rapidity correlations}
The $p_T$ integrated ($p_T>0.5\,{\rm GeV}$) event-by-event rapidity distributions $dN/d\eta_p$ are then used to determine the two-particle rapidity correlation function \cite{Vechernin:2013vpa}
\begin{equation}
  C(\eta_1,\eta_2) = \frac{\langle N(\eta_1)N(\eta_2) \rangle}{\langle N(\eta_1) \rangle \langle N(\eta_2) \rangle}\,,
\end{equation}
where $N(\eta)=dN/d\eta_p$.
We follow the ATLAS collaboration \cite{ATLAS-CONF-2015-020} and remove the effect of residual centrality dependence in the average shape $\langle N(\eta) \rangle$ by redefining the correlation function as \cite{Jia:2015jga}
\begin{equation}
  C_N(\eta_1,\eta_2) = \frac{C(\eta_1,\eta_2)}{ C_p(\eta_1) C_p(\eta_2)}\,,
\end{equation}
where 
\begin{equation}
  C_p(\eta_{1/2}) = \frac{1}{2Y}\int_{-Y}^{Y} C(\eta_1,\eta_2) d\eta_{2/1}\,.
\end{equation}
Importantly, the resulting distribution is then normalized such that the average value of $C_N(\eta_1,\eta_2)$ is one.

Following \cite{Bzdak:2012tp,Jia:2015jga} $C_N(\eta_1,\eta_2)$ is then decomposed into Legendre polynomials.
The Legendre coefficients are given by
\begin{align}
  a_{n,m} =& \int C_N(\eta_1,\eta_2) \nonumber\\ &~~ \times \frac{T_n(\eta_1) T_m(\eta_2) + T_n(\eta_2) T_m(\eta_1)}{2} \frac{d\eta_1}{Y} \frac{d\eta_2}{Y}\,,
\end{align}
where $T_n(\eta_p) = \sqrt{n+1/2}\,P_n(\eta_p/Y)$ and $P_n$ are the standard Legendre polynomials.
The $a_{n,m}$ are related to the Legendre coefficients of the single particle distribution $a_n$ via
$a_{n,m}=\langle a_n a_m\rangle$, where the $a_n$ are defined through $N(\eta)/\langle N(\eta) \rangle = 1 + \sum_n a_n T_n(\eta)$ \cite{ATLAS-CONF-2015-020}.

\section{Results}
For the calculations presented in this work, the parameters of the initial state model are adjusted to fit the measured $dN/d\eta_p$  \cite{Bearden:2001qq,Alver:2010ck,Abbas:2013bpa} and $dN/d^2p_T$ \cite{Adler:2003cb} distributions of charged hadrons, as well as the net-baryon rapidity distribution \cite{Bearden:2003hx,Arsene:2009aa}, when using the $T$-dependent bulk viscosity and $\eta/s=0.12$. In particular we use a quark-quark cross section of $3\,{\rm mb}$ and $\lambda=0.66$. The hydrodynamic evolution starts at  $\tau_0=0.38\,{\rm fm}/c$.

\begin{figure}[htb]
   \begin{center}
     \includegraphics[width=9cm]{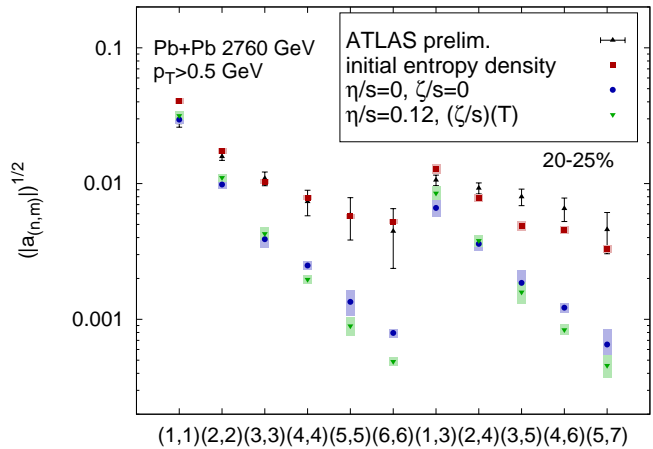}
     \caption{(Color online) Legendre coefficients $\sqrt{|a_{n,m}|}$ labeled by $(n,m)$ from the initial entropy density distribution and the final charged hadrons for 20-25\% central collisions. Hydrodynamic expansion reduces the coefficients more efficiently with increasing $n,m$. Including shear and bulk viscosity leads to an increase of the lower  $\sqrt{|a_{n,m}|}$ and a decrease of the higher $\sqrt{|a_{n,m}|}$.}
     \label{fig:anm-PbPb}
   \end{center}
   \vspace{-0.5cm}
\end{figure}

We present results for the $\sqrt{|a_{n,m}|}$ of the fluctuating initial entropy density distribution (averaged over the transverse directions) and the final $\sqrt{|a_{n,m}|}$ obtained from the produced charged hadrons with $|\eta_1|,|\eta_2|<2.4$ for 20-25\% central Pb+Pb collisions at 2760 GeV in Fig.\,\ref{fig:anm-PbPb}.
Apart from the $a_{n,n} = \langle a_n^2 \rangle$, the next largest correlations are those for $m=n+2$. For this combination of Legendre indices, the $a_{n,m}$ are negative, in qualitative agreement with the experimental data from ATLAS \cite{ATLAS-CONF-2015-020}. The $a_{n,n+1}$ vanish in symmetric collision systems. 

The final $\sqrt{|a_{n,m}|}$ show a reduction relative to those obtained from the initial entropy density distribution. This reduction increases with $n,m$, showing that the hydrodynamic expansion smears out the shorter range correlations more efficiently. The effect of viscosity is to slightly increase the lower $a_{n,m}$ and decrease the higher ones. This can be understood as the effect of diffusion that destroys shorter scale structures in $N(\eta_p)$, while the reduction of the longitudinal pressure keeps the pseudorapidity distribution more compact as shown previously \cite{Schenke:2011bn,vanderSchee:2015rta}, leading to smaller suppression of the lower $a_{n,m}$ that are sensitive to long range correlations.

Generally, the $n$ dependence of the calculated final state  $\sqrt{|a_{n,m}|}$ is stronger than observed in the experimental data \cite{ATLAS-CONF-2015-020}. The calculated $a_{n,m}$ agree with the experimental data for small $n=m=1$, but are smaller than the experimental values for all other $n,m$.

The initial state values of $\sqrt{|a_{n,m}|}$ agree fairly well with the experimental data, which is interesting, however, a direct comparison is difficult in this case.

The final calculated $\sqrt{|a_{n,m}|}$ for large $n,m$ are much smaller than the data for any choice of transport parameters. Since we are only interested in the effects of flow in this study we have not included any non-flow effects, such as short range correlations from resonance decays (see e.g. \cite{Bozek:2012en}). The large difference between the calculation and the experimental data for larger $n,m$ indicates that these have a significant effect for $n,m\geq 3$. We leave a quantitative study of this effect using statistical hadronization for future work.

\begin{figure}[htb]
   \begin{center}
     \includegraphics[width=9cm]{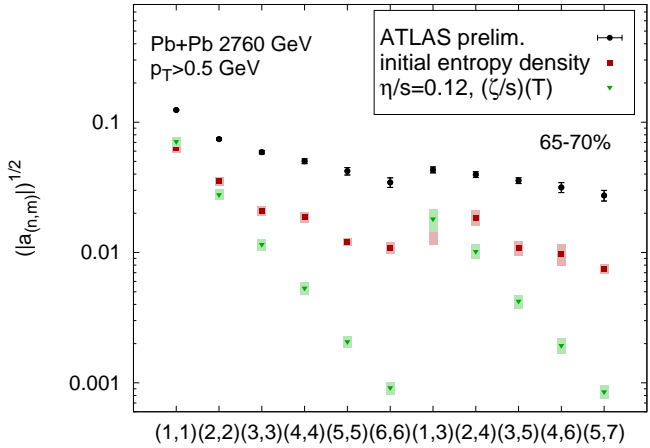}
     \caption{(Color online) Legendre coefficients $\sqrt{|a_{n,m}|}$ from the initial entropy density distribution and the final charged hadron distribution for 65-70\% central collisions.}
     \label{fig:anm-PbPb-65-70}
   \end{center}
   \vspace{-0.5cm}
\end{figure}

The same effect is seen in 65-70\% central collisions shown in Fig.\,\ref{fig:anm-PbPb-65-70}. Here the coefficients extracted from the initial distribution already underestimate the experimental data and after hydrodynamic evolution most $\sqrt{|a_{n,m}|}$ are largely underestimated. 

The agreement of the lower $a_{n,m}$ with experimental data for 20-25\% centrality, but disagreement for 65-70\%, indicates that the simple initial state model does not describe correctly the centrality dependence of the number of sources, which dictates the degree of fluctuations.

\begin{figure}[htb]
   \begin{center}
     \includegraphics[width=9cm]{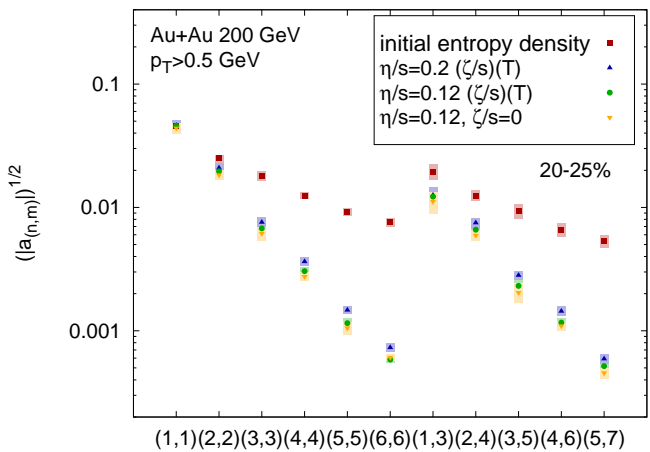}
     \caption{(Color online) Legendre coefficients $\sqrt{|a_{n,m}|}$ from the initial entropy density distribution and the final charged hadron distribution for 20-25\% central Au+Au collisions at 200 GeV. Larger shear viscosity leads to larger $\sqrt{|a_{n,m}|}$. So does the inclusion of bulk viscosity.}
     \label{fig:anm}
   \end{center}
  \vspace{-0.5cm}
\end{figure}

In Fig. \ref{fig:anm} we show predictions for $\sqrt{|a_{n,m}|}$ in Au+Au collisions at 200 GeV. We find an overall increase of all $\sqrt{|a_{n,m}|}$ compared to the Pb+Pb result at 2760 GeV by approximately $50\,\%$. We further study the effect of varying the shear viscosity to entropy density ratio $\eta/s$ from 0.12 to 0.2, and the effect of bulk viscosity separately. As opposed to the change of $\eta/s$ from 0 to 0.12 for Pb+Pb collisions (see Fig. \ref{fig:anm-PbPb}), we generally find an increase (or no change) of all $\sqrt{|a_{n,m}|}$ when increasing $\eta/s$. This indicates that the reduction of the longitudinal pressure dominates over the increased diffusion.
Including bulk viscosity has a similar effect as increasing the shear viscosity, however, the effect is not large compared to the statistical errors of our results.

\begin{figure}[htb]
   \begin{center}
     \includegraphics[width=9cm]{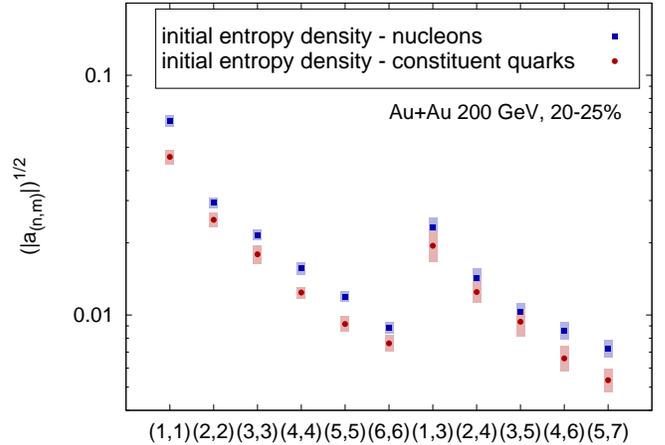}
     \caption{(Color online) Legendre coefficients $\sqrt{|a_{n,m}|}$ from the initial entropy density distribution. More sources lead to less fluctuations of the entropy density in pseudorapidity and thus smaller $a_{n,m}$.}
     \label{fig:anm-init}
   \end{center}
   \vspace{-0.5cm}
\end{figure}

Next we explicitly study the dependence of the $a_{n,m}$ on the number of sources in the initial state. Fig.\,\ref{fig:anm-init} shows how the $a_{n,m}$ of the initial entropy density distribution are increased when reducing the number of sources by employing nucleons instead of valence quarks. As alluded to above, this effect is expected, because fewer sources lead to more fluctuations.

\begin{figure}[htb]
   \begin{center}
     \includegraphics[width=9cm]{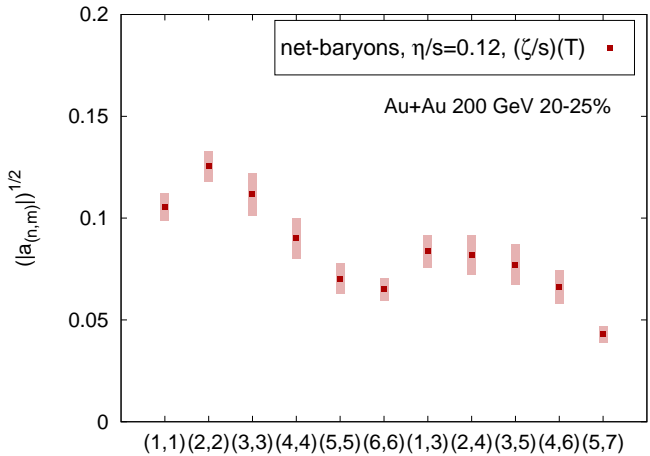}
     \caption{(Color online) Legendre coefficients $\sqrt{|a_{n,m}|}$ of the net-baryon distriution in 20-25\% central 200 GeV Au+Au collisions at RHIC. In contrast to charged hadron correlations,the second coefficient $\sqrt{|a_{(2,2)}|}=\sqrt{\langle a_2^2 \rangle}$ is larger than the first $\sqrt{|a_{(1,1)}|}$.    \label{fig:anm-netbaryon}}
   \end{center}
  \vspace{-0.5cm}
\end{figure}

Finally, in Fig.\,\ref{fig:anm-netbaryon} we present a prediction for $\sqrt{|a_{n,m}|}$ extracted from the net-baryon distribution. We see that the hydrodynamic evolution does not dampen the higher $a_{n,m}$ more than the lower ones as was the case for the charged hadron distribution. We have checked that the ideal results agree with the ones shown for $\eta/s=0.12$ and $(\zeta/s)(T)$ within errors.
We argue that measuring these quantities e.g. at RHIC and comparing to various models could reveal important details on the mechanism of baryon stopping and transport in heavy ion collisions.

\section{Conclusions and Outlook}
We have presented the first calculation of two-particle pseudorapidity correlations in a viscous fluid dynamic framework and studied the dependence of our results on the initial state and the transport properties of the medium. We have shown that the number of particle producing sources affects the Legendre coefficients $a_{n,m}$, as does the presence of hydrodynamic evolution. The shear and bulk viscosities of the medium also modify the $a_{n,m}$ coefficients. The effect of viscosity is two-fold. On the one hand, it reduces the longitudinal pressure, leading to an increase of the $a_{n,m}$. On the other hand, the diffusion has the opposite effect. This effect becomes larger for shorter range correlations, i.e., larger $n,m$.
The underestimation of the experimental $\sqrt{|a_{n,m}|}$ coefficients in particular for large $n$ and $m$ indicates that short range non-flow correlations have a large effect on these observables.

Based on our results we conclude that the study of $a_{n,m}$ together with the Fourier coefficients $v_n$ at various collision energies has the strong potential to constrain the initial state particle production mechanism as well as the transport properties of the medium.

We further propose to perform the same experimental analysis using net-baryon or net-proton distributions to constrain in detail the mechanism of baryon stopping and transport in heavy ion collisions.

 It will also be very interesting to analyze the two-particle pseudorapidity correlations using a more sophisticated initial state model, such as the IP-Glasma \cite{Schenke:2012wb,Schenke:2012hg,Gale:2012rq} in the future. This will require an extension of the boost-invariant Glasma picture to three dimensions.

\emph{Note:} An independent study of the two-particle pseudorapidity correlations within a similar framework and leading to similar conclusions has appeared simultaneously with our work \cite{Bozek:2015tca}.

{\bf Acknowledgments} We thank Gabriel Denicol, Sangyong Jeon, and Jian\-yong Jia for useful discussions.
AM is supported by the RIKEN Special Postdoctoral Researcher program. BPS is supported under DOE Contract No. DE-SC0012704. This research used resources of the National Energy Research Scientific Computing Center, which is supported by the Office of Science of the U.S. Department of Energy under Contract No. DE-AC02-05CH11231. BPS acknowledges a DOE Office of Science Early Career Award.

\vspace{-0.5cm}
\bibliography{spires}

\end{document}